\documentclass[reprint,superscriptaddress,amssymb,amsmath,aps,prapplied,longbibliography]{revtex4-1}

\usepackage{floatrow}
\usepackage{graphicx}
\usepackage{amssymb}
\usepackage{epstopdf}
\usepackage{upgreek}
\usepackage{dcolumn}
\usepackage{bm}
\usepackage{gensymb}
\usepackage{braket}
\usepackage{amsmath}
\usepackage{mathtools}
\usepackage{float}
\usepackage{tikz}
\usepackage[colorlinks=true,allcolors=blue]{hyperref}

\usepackage{array}
\newcolumntype{C}[1]{>{\centering\arraybackslash}m{#1}}

\usepackage{xcolor}

\expandafter\ifx\csname package@font\endcsname\relax\else
\expandafter\expandafter
\expandafter\usepackage
\expandafter\expandafter
\expandafter{\csname package@font\endcsname}%
\fi

\newcommand{\microwatt}{$\upmu$W}

\newcommand{\be}{\begin{eqnarray}}
\newcommand{\ee}{\end{eqnarray}}
\newcommand{\bfig}{\begin{figure}}
	\newcommand{\efig}{\end{figure}}

\DeclareFontFamily{U}{mathb}{}
\DeclareFontShape{U}{mathb}{m}{n}{
	<-5.5> mathb5
	<5.5-6.5> mathb6
	<6.5-7.5> mathb7
	<7.5-8.5> mathb8
	<8.5-9.5> mathb9
	<9.5-11.5> mathb10
	<11.5-> mathbb12
}{}

\begin{document}
	
	\title{Performance of a Kinetic-Inductance Traveling-Wave Parametric Amplifier at 4\,Kelvin: Toward an Alternative to Semiconductor Amplifiers}
	\author{M. Malnou}
	\email{maxime.malnou@nist.gov}
	\affiliation{National Institute of Standards and Technology, Boulder, Colorado 80305, USA}
	\affiliation{Department of Physics, University of Colorado, Boulder, Colorado 80309, USA}
	\author{J. Aumentado}
	\affiliation{National Institute of Standards and Technology, Boulder, Colorado 80305, USA}
	\author{M. R. Vissers}
	\affiliation{National Institute of Standards and Technology, Boulder, Colorado 80305, USA}
	\author{J. D. Wheeler}
	\affiliation{National Institute of Standards and Technology, Boulder, Colorado 80305, USA}
	\author{J. Hubmayr}
	\affiliation{National Institute of Standards and Technology, Boulder, Colorado 80305, USA}
	\author{J. N. Ullom}
	\affiliation{National Institute of Standards and Technology, Boulder, Colorado 80305, USA}
	\affiliation{Department of Physics, University of Colorado, Boulder, Colorado 80309, USA}
	\author{J. Gao}
	\affiliation{National Institute of Standards and Technology, Boulder, Colorado 80305, USA}
	\affiliation{Department of Physics, University of Colorado, Boulder, Colorado 80309, USA}
	\date{\today}
	
	\begin{abstract}
		 Most microwave readout architectures in quantum computing or sensing rely on a semiconductor amplifier at 4\,K, typically a high-electron mobility transistor (HEMT). Despite its remarkable noise performance, a conventional HEMT dissipates several milliwatts of power, posing a practical challenge to scale up the number of qubits or sensors addressed in these architectures. As an alternative, we present an amplification chain consisting of a kinetic-inductance traveling-wave parametric amplifier (KI-TWPA) placed at 4\,K, followed by a HEMT placed at 70\,K, and demonstrate a chain-added noise $T_\Sigma = 6.3\pm0.5$\,K between 3.5 and 5.5\,GHz. While, in principle, any parametric amplifier can be quantum limited even at 4\,K, in practice we find the KI-TWPA's performance limited by the temperature of its inputs, and by an excess of noise $T_\mathrm{ex} = 1.9$\,K. The dissipation of the KI-TWPA's rf pump constitutes the main power load at 4\,K and is about one percent that of a HEMT. These combined noise and power dissipation values pave the way for the KI-TWPA's use as a replacement for semiconductor amplifiers.
		 
	\end{abstract}

	\maketitle

	\section{INTRODUCTION}
	
	Superconducting parametric amplifiers have been studied and refined for decades \cite{yurke1989observation,castellanos2008amplification,bergeal2010phase,eom2012wideband,macklin2015near,frattini2017three,planat2020photonic,malnou2021three}, yet they have always been used in the same configuration: as pre-amplifiers placed at millikelvin temperatures, followed by a 4\,K stage low noise amplifier, conventionally a high-electron mobility transistor (HEMT). While Al-based parametric amplifiers can only operate well below the critical temperature of aluminum ($T_c\sim1.2$\,K), Nb-based Josephson amplifiers ($T_c\sim9$\,K) \cite{castellanos2008amplification,macklin2015near, malnou2018optimal},  or NbTiN-based kinetic amplifiers ($T_c\sim14$\,K) \cite{eom2012wideband,chaudhuri2017broadband,malnou2021three,shu2021nonlinearity,parker2021near} can operate at much higher temperatures, in particular at 4\,K.
	
	At 4\,K, HEMTs are commercially available and typically achieve input noise temperatures of a just few kevlins, with bandwidths spanning several gigahertz. They are integral to superconducting quantum computer architectures \cite{arute2019quantum}, to dark matter searches \cite{brubaker2017first,du2018search,Backes2021a}, and to the readout of superconducting transition-edge sensors or microwave kinetic inductance detectors \cite{dober2017Microwave,mcrae2020Materials}. However, cryogenic HEMTs require several milliwatts of power, and the dissipated heat load can quickly become a serious challenge when designing experiments that require scaling to massive detector or qubit channel counts. For example, in both the Lynx \cite{Bandler2019Lynx,DiPirro2019Lynx} and Origin Space Telescope \cite{Echternach2021large,bradley2021on,wiedner2021heterodyne}, two of the four large mission concepts for observatories presented in the 2020 Astronomy and Astrophysics Decadal Survey \cite{nasa2020decadal}, about 10 HEMTs are planned to measure signals from roughly 10,000 readout channels, and the power dissipation from the HEMTs is the single largest power load on the 4\,K stage of the instrument. In space, achieving 10\,mW of cooling power at 4\,K is extremely challenging, therefore techniques for measuring gigahertz signals that can reduce this heat load are of great interest.
	
	At millikelvin temperatures, the NbTiN-based kinetic-inductance traveling-wave parametric amplifier (KI-TWPA) has shown promising performance \cite{eom2012wideband,chaudhuri2017broadband,zobrist2019wide,vissers2016low,malnou2021three,shu2021nonlinearity}. In particular, its gigahertz bandwidth along with nanowatt input saturation power makes it compatible with high channel count applications, and it can operate close to the quantum limit \cite{zobrist2019wide,malnou2021three}. Furthermore, NbTiN resonators have internal quality factors $Q\gtrsim10^3$ at 4\,K due to their high $T_c$ \cite{zmuidzinas2012superconducting}, so the KI-TWPA chip should be almost dissipation-less. In addition, the three-wave mixing (3WM) mode of operation has reduced its pump power requirements to the few microwatts range \cite{ranzani2018kinetic,malnou2021three}, suggesting that the power dissipation at 4\,K can be made much smaller than that generated by common HEMTs.
	
	In this article, we ask-- can a KI-TWPA replace a conventional 4\,K stage semiconductor amplifier? Although one might expect parametric gain at 4\,K, can one also expect it to retain its noise performance at such high temperature? Here we show that, in principle, a ``hot'' parametric amplifier can be quantum-limited, as long as its input fields are well-thermalized to a cold bath. Using a shot-noise tunnel junction (SNTJ), we measure the total chain-added noise of an amplification chain configured with the KI-TWPA at 4\,K, followed by a HEMT placed at 70\,K, and find $T_\Sigma=6.3\pm0.5$\,K between 3.5 and 5.5\,GHz. It is comparable to that of a well-optimized chain with a single HEMT placed at 4\,K. Accounting for the contribution of each stage in the amplification chain, we estimate that the KI-TWPA alone generates $1.9\pm0.2$\,K of excess noise, on par with the noise added by most commercial HEMTs \cite{lnf}. Meanwhile, the power load at 4\,K, predominantly due to dissipating the rf pump, is currently about $100\,$\microwatt, or one percent that of a HEMT. Together, these measurements pave the way toward a more power efficient and lower noise alternative to semiconductor amplifiers at 4\,K.

	\section{THEORY AND EXPERIMENTAL PRINCIPLE}
	\label{sec:theo}
	
	The fundamental limit on the noise added by a lossless, phase-insensitive parametric amplifier is often described as originating from an internal mode \cite{caves1982quantum,Epstein2021quantum,yamamoto2003noise}. But this mode is not an inaccessible internal degree of freedom. Rather, it refers to the input ``idler'' mode, and therefore this ideal amplifier should properly be described as a 4-port device: two inputs and two outputs, at the signal and idler frequencies (the idler output is usually not monitored). In this context, the mean amplifier output signal power is, in units of photons \cite{malnou2021three}:
	\begin{equation}
	    N_\mathrm{out}^s = G N_\mathrm{in}^s + (G-1)N_\mathrm{in}^i,
	\label{eq:Nouts}
 	\end{equation}
	where $G$ is the amplifier signal power gain and $N_\mathrm{in}^s$ ($N_\mathrm{in}^i$) is the input power at the signal (idler) frequency. Here, regardless of its physical temperature, the quantum-limited nature of the amplifier only depends on the input idler state: when it is vacuum $N_\mathrm{in}^i=1/2$, in the high gain limit the amplifier adds half a photon worth of noise energy to the input-referred signal $N_\mathrm{in}^s$.
	
\begin{figure}[h] 
	\centering
	\includegraphics[scale=0.49]{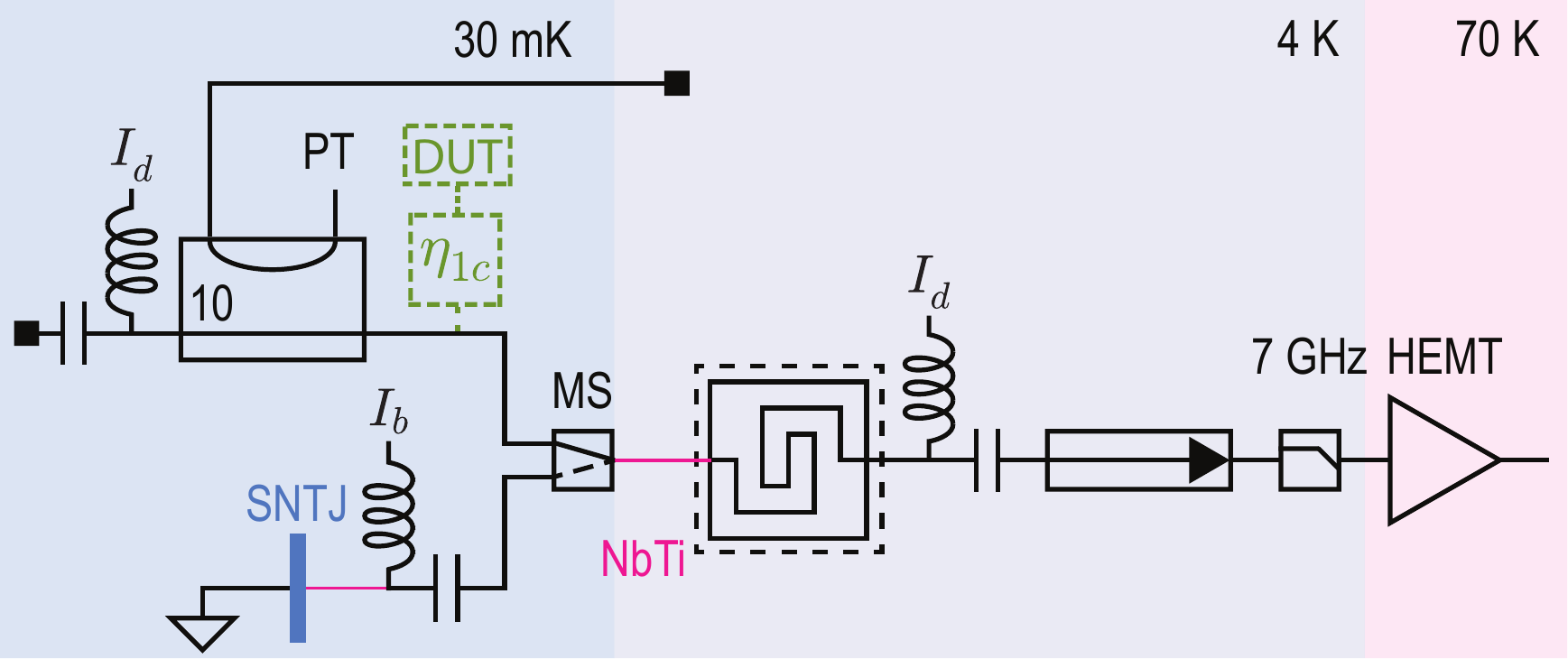}
	\caption{Schematic of the amplification chain. The KI-TWPA (squared spiral) is magnetically shielded (dashed square) and placed at 4\,K while the HEMT operates at 70\,K, protected from the strong KI-TWPA pump tone (PT) by a low pass filter. The PT then dissipates into the isolator (4-12\,GHz). When the microwave switch (MS) is in its top position, the PT is delivered to the KI-TWPA via a directional coupler (DC), while two bias tees (BT) allow the dc current $I_d$ to flow in and out of the chain, through the KI-TWPA. A hypothetical DUT is coupled to the chain with efficiency $\eta_\mathrm{1c}$. When the MS is in its bottom position, the SNTJ delivers a known noise \cite{spietz2006shot} to the chain's input, whose power depends on the current $I_b$ with which the SNTJ is biased.}
    \label{fig:setup}
    \end{figure}

    To test whether a ``hot'' parametric amplifier can remain quantum-limited, we design the amplification chain illustrated in Fig.\,\ref{fig:setup}, where a KI-TWPA is placed at 4\,K, but whose signal and idler inputs are connected to the millikelvin stage. Then, provided that (i) the KI-TWPA's gain is enough overcome the following loss and amplifier-added noise, (ii) the KI-TWPA's inputs are cold, and (iii) the KI-TWPA is lossless, i.e. does not add any excess of noise on top of that from its inputs, the noise added by the entire chain should reach the quantum limit (half a photon), see appendix\,\ref{app:chainnoise}.
    
    Operating the KI-TWPA in a 3WM fashion \cite{vissers2016low,malnou2021three}, we employ a bias tee (BT) and a directional coupler (DC) to deliver respectively a dc current and an rf pump to its physical input port. Both BT and DC are placed at 30\,mK, behind a hypothetical device under test (DUT) coupled to the readout line thereby minimizing insertion loss between DUT and KI-TWPA. This configuration is particularly suitable when the DUT represents an array of resonators, such as microwave kinetic-inductance detectors \cite{szypryt2017large,zobrist2019wide}, or a microwave superconducting quantum interference device multiplexer \cite{Irwin2004microwave,mates2017simultaneous}.
    
    \begin{figure*}[!t] 
	\centering
	\includegraphics[scale=0.6]{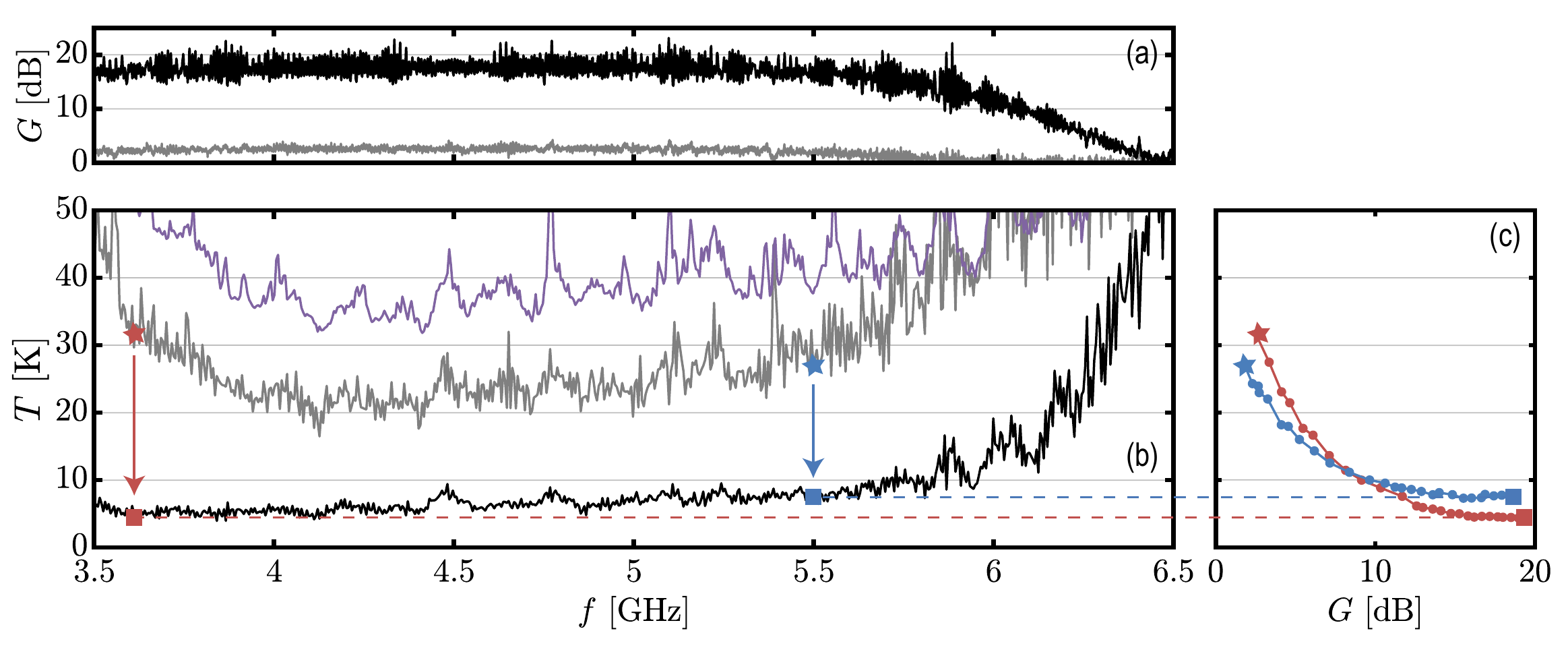}
	\caption{Chain-added noise measurement. (a) The KI-TWPA gain $G$ (ratio pump on/off) as a function of frequency (measured with a vector network analyzer) is shown for two rf pump powers ($P_p$): at low pump power (gray curve, $P_p=-37$\,dBm at the KI-TWPA's input) and high pump power (black curve, $P_p=-29$\,dBm). In both cases, the pump frequency is $f_p=8.979$\,GHz, and the KI-TWPA is dc biased with $I_d=0.7$\,mA. (b) We first measured the frequency-dependent chain-added noise $T_\Sigma'$ (purple curve) with the SNTJ. Below 4\,GHz and above 6\,GHz, $T_\Sigma'$ increases because we reach the pass-band edge of the components in the amplification chain (HEMT, isolator and low-pass filter). Then, we deduced the chain-added noise $T_\Sigma$ when the KI-TWPA operates at the low (gray curve) and high (black curve) gain $G$ presented in panel (a). Points at two witness frequencies (3.6\,GHz and 5.5\,GHz) underline how the noise diminishes (from stars to squares) when the gain increases. (c) At these two frequencies, the variation of $T_\Sigma$ as a function of gain shows that the lowest noise temperature (square) reaches an asymptote.
	} 
    \label{fig:NVR}
    \end{figure*}

	To measure the added noise of such a chain, we insert a microwave switch (MS) that allows us to alternate between the KI-TWPA biasing components and a calibrated noise source, consisting of a shot noise tunnel junction (SNTJ) \cite{Spietz2003primary,malnou2021three}. We first obtain the added noise $N_\Sigma'$ of the chain when the MS is toggled toward the SNTJ (see appendix\,\ref{app:fitshot}). Then, after actuating the MS, we turn on the KI-TWPA and measure the level $r$ to which the output noise rises. We then deduce the chain-added noise $N_\Sigma$ using
	\begin{equation}
	    \frac{r}{G} = \frac{N_\Sigma + N_c} {N_\Sigma' + N_c},
	\label{eq:r}
	\end{equation}
	where $G$ is the KI-TWPA gain, see appendix\,\ref{app:noise rise}.
    
    \section{PERFORMANCE AT 4\,K}
	\label{sec:per4K}
	
	The results of such a measurement are presented in Fig.\,\ref{fig:NVR}, performed when using a commercial HEMT and a KI-TWPA, whose design and millikelvin performance have been described elsewhere \cite{malnou2021three}. With the MS actuated toward the DC and BT, we operate the KI-TWPA at various gains, from an average of $2.5$\,dB, to $18$\,dB between 3.5 and 5.5 GHz, see Fig.\,\ref{fig:NVR}(a); the higher gain profile remains flat, with less than $3$\,dB ripples in that band. For each operating gain $G$, we record the noise rise $r$ on a spectrum analyzer (SA) and form $T_\Sigma=N_\Sigma\hbar\omega/k_B$ (with $\hbar$ the reduced Planck constant and $k_B$ the Boltzmann constant) using Eq.\,\ref{eq:r}. In Fig.\,\ref{fig:NVR}(b), we show $T_\Sigma$ as a function of frequency, when the KI-TWPA is operated at low and high gain (gray and black curves, respectively), along with $T_\Sigma'=N_\Sigma'\hbar\omega/k_B$, obtained when the MS is actuated toward the SNTJ (purple curve). At high gain, $T_\Sigma=6.3\pm0.5$\,K between $3.5$ and $5.5$\,GHz (with uncertainty dominated by that of the chain's output power, known within $\pm$0.3\,dB). To our knowledge, it is the first time that such a low and broadband noise performance has been obtained without the use of a semiconductor amplifier at 4\,K.
	
	As discussed in Sec.\,\ref{sec:theo}, three possible sources of noise prevent $T_\Sigma$ to reach the quantum limit: (i) insufficient KI-TWPA gain, (ii) warm KI-TWPA inputs, and (iii) KI-TWPA-excess noise. Although from a user perspective only $T_\Sigma$ matters, knowing the separate contribution of these sources is interesting from an amplifier-design perspective, because it indicates if and how $T_\Sigma$ may be improved. In Fig.\,\ref{fig:NVR}(c) we present the variation of $T_\Sigma$ as a function of $G$, at two different frequencies. As $G$ increases, $T_\Sigma$ reaches an asymptotic lower bound. Therefore, the high gain regime of our KI-TWPA is enough to overcome the following loss and HEMT-added noise; operating at higher gain would not improve the chain-added noise.
	
	To evaluate the noise coming from the KI-TWPA inputs, we separately measure at 4\,K the transmission efficiencies of each stage in the amplification chain, see appendix\,\ref{app:lossbudg}. Each efficiency acts as an effective noise source, whose temperature is governed by a beamsplitter interaction (see appendix\,\ref{app:chainnoise}). In Tab.\,\ref{tab:lossbudg} we report the transmission efficiencies and corresponding noise temperatures of the various stages, either when considered separately (intrinsic noise) or within the amplification chain (chain-input-referred noise). Clearly, the temperature of the KI-TWPA inputs is dominated by the effect of $\eta_\mathrm{1h}=0.8$, the transmission efficiency at 4\,K between the DUT and the KI-TWPA. It generates $2.1$\,K of noise at the input of the amplifier (and $2.6$\,K when referred to the input of the chain). Practically, $T_\Sigma$ may be significantly decreased with increasing $\eta_\mathrm{1h}$: for example with $\eta_\mathrm{1h}=0.9$, $T_\Sigma$ would drop from $6.3$\,K to $4.5$\,K. We believe this performance achievable, because most of our warm insertion loss originates in the KI-TWPA packaging, whose printed-circuit boards and connectors may be made less dissipative.

    Subtracting the (input-referred) noise generated by all the inefficient transmissions from $T_\Sigma$, we deduce the KI-TWPA-excess noise: at the chain's input, it amounts to 2.9\,K, equivalent to an intrinsic excess noise temperature $T_\mathrm{ex}=1.9\pm0.2$\,K, on par with that of the HEMT at 4\,K \cite{lnf}. We are currently investigating the origin of this noise, which might come from quasi-particle generation.
	
	\begin{table}[!h] 
        \centering
        \begin{tabular}{c||C{0.63cm}C{0.63cm}C{0.63cm}C{0.63cm}C{0.63cm}}
        \textbf{\begin{tabular}[c]{@{}c@{}}sources of\\ noise\end{tabular}} &
        \includegraphics[scale=0.7]{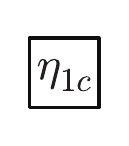} & \includegraphics[scale=0.7]{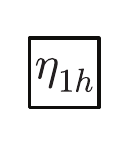} & \includegraphics[scale=0.7]{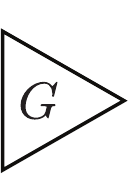} & \includegraphics[scale=0.7]{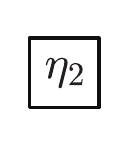} & \includegraphics[scale=0.7]{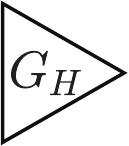}  \\ \hline
         \textbf{\begin{tabular}[c]{@{}c@{}}transmission\\ efficiency\end{tabular}} & 0.80 & 0.80 & - & 0.61 & -\\ \hline
        \textbf{\begin{tabular}[c]{@{}c@{}}insertion\\ loss {[}dB{]}\end{tabular}} & 1 & 1 & - & 2.2 & -\\ \hline
        \textbf{\begin{tabular}[c]{@{}c@{}}intrinsic \\ noise {[}K{]}\end{tabular}} & 0.16 & 2.1 & 1.9 & 2.7 & 13.4 \\ \hline
        \textbf{\begin{tabular}[c]{@{}c@{}}input-referred\\ noise {[}K{]}\end{tabular}} & 0.16 & 2.6 & 2.9 & 0.07 & 0.6 \\
        \end{tabular} 
        \caption{Noise contribution of each stage of the amplification chain, averaged between 3.5 and 5.5\,GHz.  DUT signals are routed to the amplifier's signal input with efficiency $\eta_\mathrm{1c}$ at 30\,mK (i.e. there is $-10\log(\eta_\mathrm{1c})$\,dB of insertion loss) and with efficiency $\eta_\mathrm{1h}$ at 4\,K. After the KI-TWPA, signals are routed with efficiency $\eta_2$ to the HEMT input, see appendix\,\ref{app:chainnoise}. From the transmission efficiencies, we calculate the intrinsic and chain-input-referred noise temperatures, using the expressions reported in Tab.\,\ref{tab:exprnoise}. Note that for the noises related to $\eta_\mathrm{1c}$ and $\eta_\mathrm{1h}$, the contribution of the signal and idler paths have been added. The intrinsic HEMT-added noise at 70\,K (first stage of our dilution refrigerator) $T_H=13.4\pm0.4$\,K lies between the HEMT-added noise at 4\,K and that at 296\,K \cite{lnf}.}
        \label{tab:lossbudg}
    \end{table}
	
	As a fair comparison to $T_\Sigma$, we measured the noise temperature $T_{\Sigma 2}$ added by a well-optimized chain, employing only a HEMT at 4\,K (see appendix\,\ref{app:4KHEMT}). Between $3.5$ and $5.5$\,GHz, $T_{\Sigma2}=3.5\pm0.3$\,K (with uncertainty here dominated by that of the SNTJ impedance, $48.2\pm3.5\,\Omega$). Therefore, while not surpassing it, the KI-TWPA-based solution approaches the HEMT-based performance.
    
    \section{POWER CONSUMPTION}
	\label{sec:pow}

    In combination with having a competitive amplification and noise performance compared to that of a HEMT, the 4\,K stage KI-TWPA is expected to consume much less power. In fact, it typically requires an rf pump power $P_p\sim-30$\,dBm (i.e. $1\,$\microwatt), several orders of magnitude lower than the power requirement for a standard ($\sim 10\,$mW \cite{lnf}) or even state-of-the-art ($300\,$\microwatt\,\cite{Cha2020A300}) HEMT. Note however that $P_p$ does not account for the dissipation along the pump line. In our current setup, the pump travels to the KI-TWPA via a 10\,dB attenuator at 4\,K, and through the weakly coupled port of a 10\,dB DC at the  millikelvin stage, see appendix\,\ref{app:expsetup}. Therefore in our experiment, a typical $-30$\,dBm pump tone delivered to the KI-TWPA translates into dissipating $99\,$\microwatt\,at 4\,K (the DC being terminated at 4\,K). But such a heavy attenuation is not mandatory: in principle, the high-pass filter (HPF) we employ at the millikelvin stage suffices to prevent the room temperature 300\,K noise to directly contaminate the signal and idler bands. At the pump frequency, this noise is negligible compared to the tone's power and therefore does not affect the dynamics of the KI-TWPA (in the limit of non-diverging gain \cite{parker2021near}). So instead of being injected through the 4\,K stage attenuator and millikelvin stage DC, the pump could simply pass through the HPF and enter the rf port of the input bias tee. Furthermore, at 4\,K, $P_p$ is currently dissipated in the isolator placed before the HEMT, which could also be avoided by using a diplexer to redirect the pump to higher temperature. With these strategies, the dissipation associated with the pump can be reduced below $1\,$\microwatt\,at 4\,K.
    
\begin{figure}[h!]
	\centering
	\includegraphics[scale=0.4]{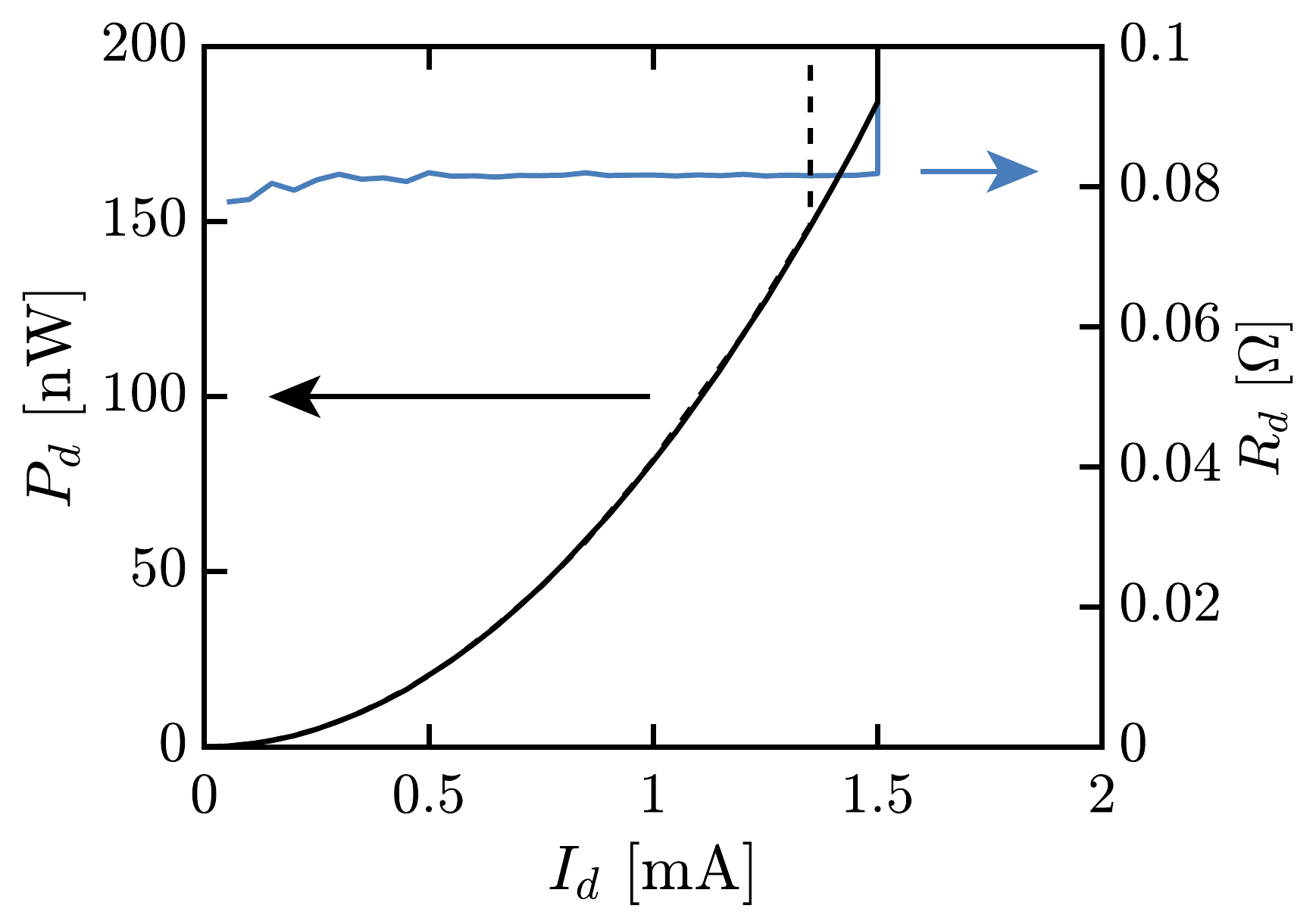} 
	\caption{Characterization of the KI-TWPA dc power dissipation. The dissipated power $P_d$ (left y-axis) is shown as a function of the biasing current $I_d$ for two situations: when the KI-TWPA's pump is off (solid black line) and when $P_p=-20$\,dBm (dashed black line). In the pump-off situation, the resistance $R_d$ (right y-axis) is also shown as a function of $I_d$ (blue curve).}
    \label{fig:pow}
    \end{figure}

    But the KI-TWPA also requires a non-negligible dc bias $I_d\sim1$\,mA, which may dissipate and generate heat. We measured the corresponding dissipated power $P_d$ with a four-point probe setup (see appendix\,\ref{app:pow}): reading the voltage drop $V_d$ across the KI-TWPA, we then retrieve $P_d = V_dI_d$. In Fig.\,\ref{fig:pow} we show $P_d$ as function of $I_d$, when the rf pump is off (solid black line) or instead, very strong ($P_p=-20$\,dBm, dashed black line). In both cases, at $I_d=1$\,mA we obtain $P_d\simeq100$\,nW, several orders of magnitude lower than what a HEMT consumes. We also show the resistance $R_d = V_d/I_d$ (right y-axis) as a function of $I_d$. Under normal operation, i.e. for $I_d<1.5$\,mA, the resistance across the four-point probes (encompassing the KI-TWPA, its packaging, and the BT at 4\,K) is $R_d=80\,\mathrm{m}\Omega$. Conversely, $R_d$ sharply increases to $\sim1\,\mathrm{k}\Omega$ when $I_d>1.5$\,mA, because superconductivity breaks down inside the KI-TWPA, probably at a weak link \cite{bockstiegel2014development,malnou2021three}. Note that the transition to this dissipative regime happens at slightly reduced $I_d$ when $P_p=-20$\,dBm, suggesting that both the dc and rf currents can activate the weak link. Voltage biasing the KI-TWPA through a small ($\sim100\,\Omega$) shunting resistor would alleviate transitioning to this regime.
    
	\section{CONCLUSION}
	
	Despite their remarkable noise performance at 4\,K, semiconductor amplifiers remain power hungry. This dissipation may limit the scale of future applications, for which large arrays ($>10^5$) of detectors or qubits would require tens to hundreds of low noise amplifiers. As an alternative, we investigated the use of a parametric amplifier at 4\,K: the KI-TWPA. Using a SNTJ as a calibrated noise source, we measured the noise added by an amplification chain where the KI-TWPA is the sole 4\,K stage amplifier. In principle, this chain can remain quantum limited; in practice, when the KI-TWPA is operated at an average gain of 18\,dB within a 2\,GHz bandwidth (and with less than 3\,dB gain ripples in that band) we measured an average chain-added noise of $6.3\pm0.5$\,K, comparable to that of a chain where the HEMT is the 4\,K stage amplifier. This performance is limited mostly by the insertion loss at 4\,K preceding the KI-TWPA, and by an excess of noise $T_\mathrm{ex}=1.9\pm0.2$\,K. Furthermore, the heat load at 4\,K, currently in the 100\,\microwatt\, range, is due to dissipating the rf pump. It is two orders of magnitude lower than what a conventional HEMT generates, and can straightforwardly be reduced below 1\,\microwatt. To our knowledge this work is the first successful implementation of a broadband, high-gain, low noise, and power-efficient microwave parametric amplifier at 4\,K. As such, our work constitutes a paradigm shift in the readout architecture for large numbers of microwave resonators.
	
	\section*{Acknowledgments} 
	
	Certain commercial materials and equipment are identified in this paper to foster understanding. Such identification does not imply recommendation or endorsement by the National Institute of Standards and Technology, nor does it imply that the materials or equipment identified are necessarily the best available for the purpose. We gratefully acknowledge support from the NIST Program on Scalable Superconducting Computing, NASA under Grant No. NNH18ZDA001N-APRA, and the DOE Accelerator and Detector Research Program under Grant No. 89243020SSC000058.
	
	\appendix
	
	\section{Added noise}
	
	\subsection{Chain-added noise}
	\label{app:chainnoise}
	
\begin{figure}[h!]
	\centering
	\includegraphics[scale=0.49]{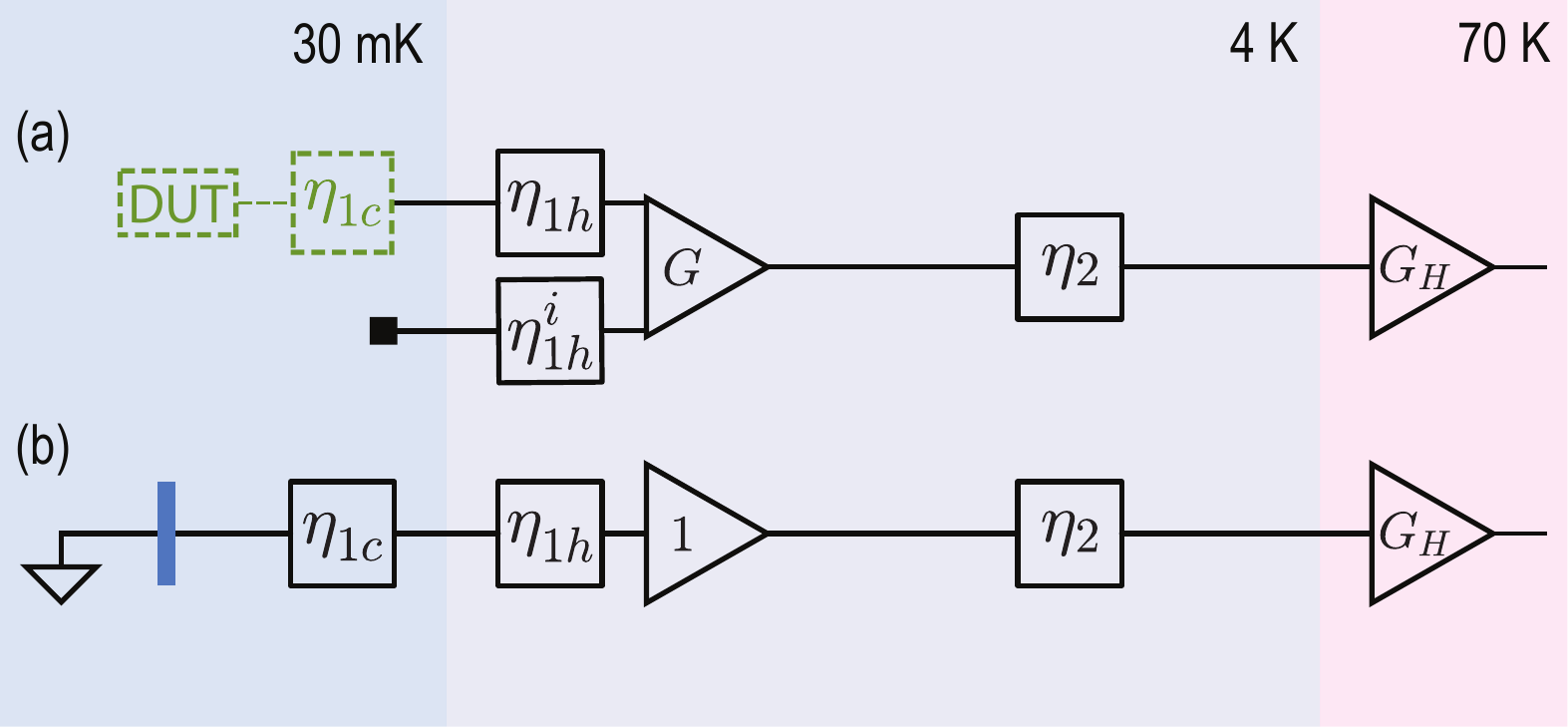} 
	\caption{Diagram of cascaded transmission efficiencies and gains, when the MS is activated toward the KI-TWPA biasing components (a), or toward the SNTJ (b).}
    \label{fig:diagr}
    \end{figure}

	 Figures \ref{fig:diagr}(a) and \ref{fig:diagr}(b) recast the amplification chain presented in Fig.\,\ref{fig:setup} in a diagram of cascaded transmission efficiencies and gains, when the microwave switch (MS) is activated toward the KI-TWPA biasing components [Fig.\,\ref{fig:diagr}(a)], or toward the SNTJ [Fig.\,\ref{fig:diagr}(b)]. Considering the situation in Fig.\,\ref{fig:diagr}(a) first, a signal with photon number $N_\mathrm{in}^s$ at the chain's input (that would be generated by the hypothetical DUT), undergoes amplification and loss when propagating to the chain:
    \begin{align}
    N_\mathrm{1c}^s &= \eta_\mathrm{1c}\left(N_\mathrm{in}^s + \frac{1-\eta_\mathrm{1c}}{\eta_\mathrm{1c}}N_c\right) \label{eq:N1cs}\\ 
    N_\mathrm{1h}^s &= \eta_\mathrm{1h}\left(N_\mathrm{1c}^s + \frac{1-\eta_\mathrm{1h}}{\eta_\mathrm{1h}}N_h\right)\\
    N_\mathrm{1h}^i &= \eta_\mathrm{1h}^i\left(N_c + \frac{1-\eta_\mathrm{1h}^i}{\eta_\mathrm{1h}^i}N_h\right)\\
    N_2^s &= G(N_\mathrm{1h}^s + N_\mathrm{ex}^s) + (G-1)(N_\mathrm{1h}^i + N_\mathrm{ex}^i)\\
    N_3^s &= \eta_2\left(N_2^s + \frac{1-\eta_2}{\eta_2}N_h\right)\\
    N_4^s &= G_H(N_3^s + N_H), \label{eq:N4ssimple}
    \end{align}    
	where the variables are all defined in Tab\,\ref{tab:varnames}. 
	
	\begin{table}[!h] 
        \centering
        \begin{tabular}{l||l}
        \hline \hline
        \textbf{\begin{tabular}[c]{@{}c@{}}Variable\\ name\end{tabular}} & \textbf{Definition} \\ \hline
        $N_\mathrm{in}^s$ & chain's input signal\\ 
        $N_c$ & Vacuum noise \\
        $N_\mathrm{1c}^s$ & Cold stage output signal \\
        $N_h$ & 4\,K stage thermal noise \\
        $N_\mathrm{1h}^s$ & KI-TWPA input signal \\
        $N_\mathrm{1h}^i$ & KI-TWPA input idler \\
        $N_\mathrm{ex}^s$ & Signal-to-signal path KI-TWPA-excess noise \\
        $N_\mathrm{ex}^i$ & Idler-to-signal path KI-TWPA-excess noise \\
        $N_\mathrm{ex}$ & overall KI-TWPA-excess noise \\
        $N_H$ & HEMT-added noise \\
        $N_2^s$ & KI-TWPA output signal \\
        $N_3^s$ & HEMT input signal \\
        $N_4^s$ & HEMT output signal \\
        $\eta_\mathrm{1c}$ & cold stage signal transmission efficiency \\
        $\eta_\mathrm{1h}$ & 4\,K stage signal transmission efficiency \\
        $\eta_\mathrm{1h}^i$ & 4\,K stage idler transmission efficiency \\
        $\eta_2$ & KI-TWPA to HEMT transmission efficiency \\
        $G$ & KI-TWPA signal power gain\\
        $G_H$ & HEMT signal power gain\\ 
        \hline \hline
        \end{tabular} 
        \caption{List of the variables pertaining to the amplification chain, Figs.\,\ref{fig:setup}b and c. The variables designating a power (named with $N$) are in units of quanta. The transmission efficiencies are dimensionless, and the gains are linear.}
        \label{tab:varnames}
    \end{table}
    
   Assuming that the HEMT gain is sufficient to overcome any following loss and amplifier-added noise, the power at the signal frequency reaching the spectrum analyzer (SA) is directly proportional to $N_4^s$. Using Eqs.\,\ref{eq:N1cs} to \ref{eq:N4ssimple}, and in the simpler case where $G\gg1$ and where idler and signal transmission efficiencies at 4\,K are equal, $\eta_\mathrm{1h}^i=\eta_\mathrm{1h}$, we have
	\begin{equation}
	  N_4^s = G_H\eta_2G\eta_\mathrm{1h}\eta_\mathrm{1c}(N_\mathrm{in}^s + N_\Sigma),
	\label{eq:N4s}
	\end{equation}
	where
	\begin{equation}
	\begin{aligned}
	  N_\Sigma = \frac{2-\eta_\mathrm{1c}}{\eta_\mathrm{1c}}N_c &+ 2 \frac{1-\eta_\mathrm{1h}}{\eta_\mathrm{1h}\eta_\mathrm{1c}}N_h + \frac{1}{\eta_\mathrm{1h}\eta_\mathrm{1c}}N_\mathrm{ex} \\ + \frac{1-\eta_2}{\eta_2G\eta_\mathrm{1h}\eta_\mathrm{1c}}N_h &+ \frac{1}{\eta_2G\eta_\mathrm{1h}\eta_\mathrm{1c}}N_H,
	\end{aligned}
	\label{eq:fullNsig}
	\end{equation}
	is the chain-added noise. The first line on the right-hand side of Eq.\,\ref{eq:fullNsig} can be identified to the KI-TWPA-added noise; it depends not only on the KI-TWPA-excess noise $N_\mathrm{ex}=N_\mathrm{ex}^s+N_\mathrm{ex}^i$, but also on the cold and warm transmission efficiencies between the KI-TWPA and the chain's input. The second line represents the contributions of the elements placed after the KI-TWPA: transmission efficiency $\eta_2$ (first term) and HEMT-added noise (second term); they are damped at sufficiently high gain $G$. In the high gain limit, with perfect transmission efficiencies $\eta_\mathrm{1c}=\eta_\mathrm{1h}=1$ (i.e. the KI-TWPA's inputs are perfectly thermalized to the cold bath) and without excess noise ($N_\mathrm{ex}=0$), we verify that $N_\Sigma=1/2$, the minimum chain-added noise.
	
	All the terms in Eq.\,\ref{eq:fullNsig} are referred to the chain's input: the noise from each source is divided by the transmission efficiencies and gain preceding it. Conversely, the noise intrinsic to each stage is chain-independent. The expressions for the chain-input-referred and intrinsic noise for each stage is reported in Tab\,\ref{tab:exprnoise}.
	
	\begin{table}[!h] 
        \centering
        \begin{tabular}{c c c}
        \textbf{\begin{tabular}[c]{@{}c@{}}sources of\\ noise\end{tabular}} & \textbf{\begin{tabular}[c]{@{}c@{}}chain-input-referred\\ noise\end{tabular}} & \textbf{\begin{tabular}[c]{@{}c@{}}intrinsic\\ noise\end{tabular}} \\
        \hline\hline
        \includegraphics[scale=0.7]{eta_1c.pdf} & $\frac{2-\eta_\mathrm{1c}}{\eta_\mathrm{1c}}N_c$ & $\frac{2-\eta_\mathrm{1c}}{\eta_\mathrm{1c}}N_c$ \\ \includegraphics[scale=0.7]{eta_1h.pdf} & 2 $\frac{1-\eta_\mathrm{1h}}{\eta_\mathrm{1h}\eta_\mathrm{1c}}N_h$ & $2\frac{1-\eta_\mathrm{1h}}{\eta_\mathrm{1h}}N_h$ \\
        \includegraphics[scale=0.7]{G.pdf} & $\frac{1}{\eta_\mathrm{1h}\eta_\mathrm{1c}}N_\mathrm{ex}$ &  $N_\mathrm{ex}$ \\
        \includegraphics[scale=0.7]{eta_2.pdf} & $\frac{1-\eta_2}{\eta_2G\eta_\mathrm{1h}\eta_\mathrm{1c}}N_h$ & $\frac{1-\eta_2}{\eta_2}N_h$ \\
        \includegraphics[scale=0.7]{GH.pdf} & $\frac{1}{\eta_2G\eta_\mathrm{1h}\eta_\mathrm{1c}}N_H$ &$N_H$ \\ \hline
        \end{tabular} 
        \caption{Noise generated by each stage in the amplification chain.}
        \label{tab:exprnoise}
    \end{table}
	
	To calculate $N_\Sigma'$ (the chain-added noise when the KI-TWPA is off and the MS actuated toward the SNTJ), we propagate the SNTJ calibrated noise $N_\mathrm{in}^s$ (which is our 'signal') through the chain shown in Fig.\,\ref{fig:diagr}(b). Here, $\eta_\mathrm{1c}$ accounts for the cold loss coming in particular from the SNTJ packaging and from the following BT. Meanwhile, the unpumped KI-TWPA acts as a passive element of gain 1, therefore we need not keep track of the noise entering the idler port  \cite{malnou2021three,ranadive2021reversed}, because the KI-TWPA output signal does not contain the idler component, as can be seen from Eq.\,\ref{eq:Nouts}. We thus obtain the HEMT-output signal power
	\begin{equation}
	  N_4^{s'} = G_H\eta_2\eta_\mathrm{1h}\eta_\mathrm{1c}(N_\mathrm{in}^s + N_\Sigma'),
	\label{eq:N4sp}
	\end{equation}
	with
	\begin{equation}
	\begin{aligned}
	  N_\Sigma' = \frac{1-\eta_\mathrm{1c}}{\eta_\mathrm{1c}}N_c &+ \frac{1-\eta_\mathrm{1h}}{\eta_\mathrm{1h}\eta_\mathrm{1c}}N_h \\ + \frac{1-\eta_2}{\eta_2\eta_\mathrm{1h}\eta_\mathrm{1c}}N_h &+ \frac{1}{\eta_2\eta_\mathrm{1h}\eta_\mathrm{1c}}N_H.
	\label{eq:Nsigmap}
	\end{aligned}
	\end{equation}
	Varying $N_\mathrm{in}^s$, we retrieve $N_\Sigma'$. Then, knowing the transmission efficiencies from the loss budget (see appendix\,\ref{app:lossbudg}), we can calculate $N_H$.

	\subsection{Noise rise}
	\label{app:noise rise}
	
	The noise rise measurement consists of comparing the output noise power, recorded on the SA, when the KI-TWPA is on and off; knowing $G$ and $N_\Sigma'$, we can retrieve $N_\Sigma$.
	
	With the MS actuated toward the KI-TWPA, the HEMT-output noise obtained with the KI-TWPA off (pump off, no dc bias) is equal to that of Eq.\,\ref{eq:N4sp} (assuming the loss between the two MS paths equal, and that signals from the DUT would be transmitted with efficiency $\eta_\mathrm{1c}$ at 30\,mK). But here, $N_\mathrm{in}^s=N_c$, because there is vacuum noise at the chain's signal input. Similarly, the HEMT-output noise obtained with the KI-TWPA on is equal to that of Eq.\,\ref{eq:N4s}, with $N_\mathrm{in}^s=N_c$, such that the ratio of output noises $r$ is
	\begin{equation}
	    r = G \frac{N_\Sigma + N_c}{N_\Sigma' + N_c},
	\end{equation}
	which gives Eq.\,\ref{eq:r}.

	\section{Full experimental setup}
	\label{app:expsetup}
	
\begin{figure*}[!t] 
	\centering
	\includegraphics[scale=0.6]{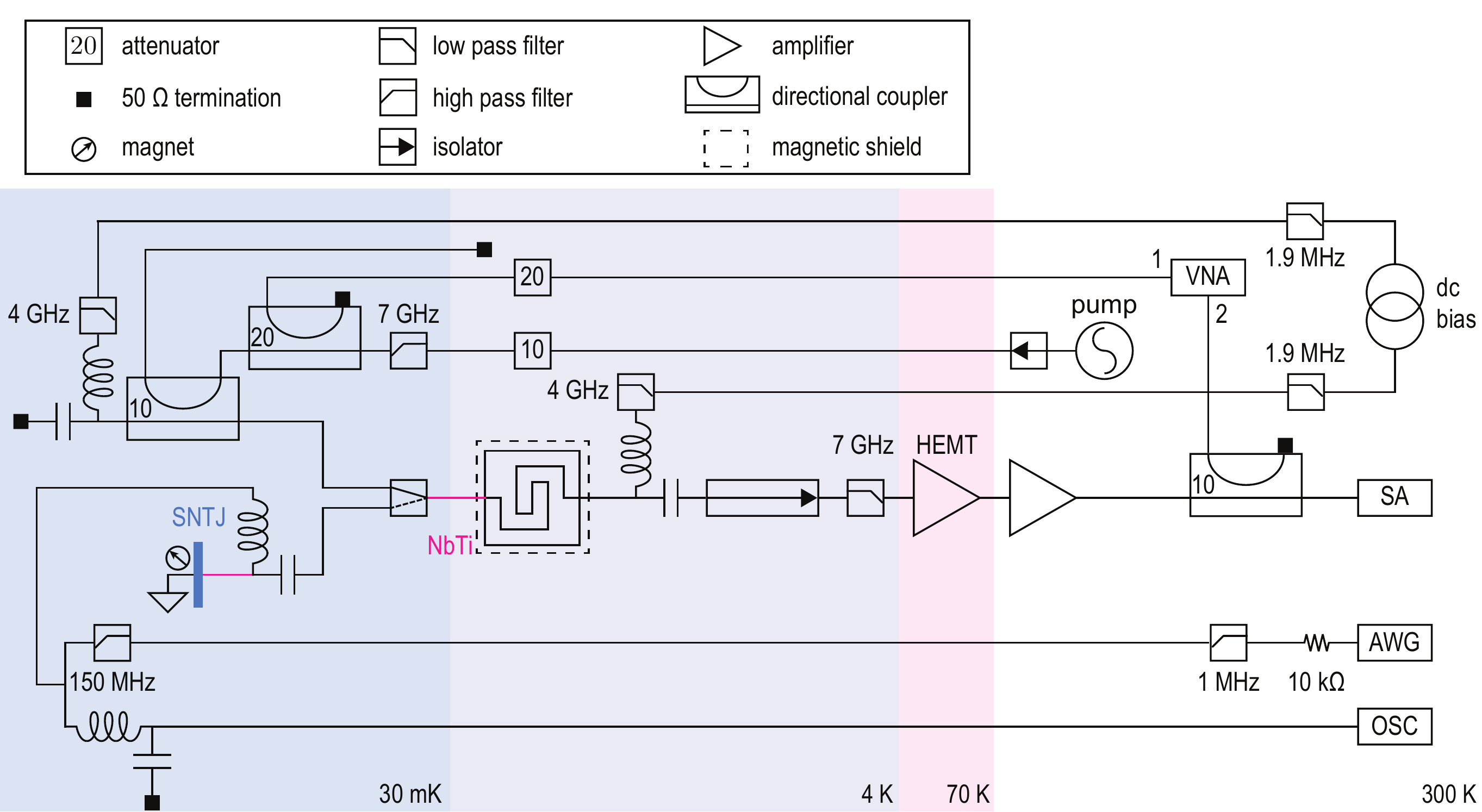}
	\caption{Full experimental setup used to measure the KI-TWPA's noise performance at 4\,K.} 
    \label{fig:fullsetup}
    \end{figure*}

	The full experimental setup used to measure the chain-added noise containing the KI-TWPA is represented in Fig.\,\ref{fig:fullsetup}. It is composed of three main parts: the KI-TWPA control electronics (top), the amplification chain (middle) and the SNTJ control electronics (bottom). At the top, a current source and a microwave generator output respectively the dc and rf KI-TWPA biases. In addition, a vector network analyzer (VNA) is connected to the input and output of the amplification chain, allowing to measure the KI-TWPA gain.
	
	In the middle, the amplification chain contains three amplifiers: the KI-TWPA at 4\,K, the HEMT at 70\,K, and a room temperature amplifier (LNA-30-00101200-17-10P). Signals from the millikelvin stage, either vacuum noise or calibrated noise from the SNTJ, depending on the microwave switch position, travel through these amplifiers and are read on the spectrum analyzer at room temperature.
	
	Because of cable resistance between the SNTJ and its drive generator (the AWG), we cannot directly voltage bias the SNTJ. Instead, we current bias it with $I_b$ (through a $10\,\mathrm{k}\Omega$ polarization resistor) and retrieve $V=R_\mathrm{SNTJ} I_b$ by first measuring the SNTJ resistance $R_\mathrm{SNTJ}$. To do that, we send a known dc current with the AWG and measure the dc voltage across the SNTJ with the oscilloscope (OSC), mounted in a 4-point probe configuration (and set with the $1\,\mathrm{M}\Omega$ input impedance).
	
	\section{4\,K HEMT amplification chain}
	\label{app:4KHEMT}
	
\begin{figure}[!h]
	\centering
	\includegraphics[scale=0.53]{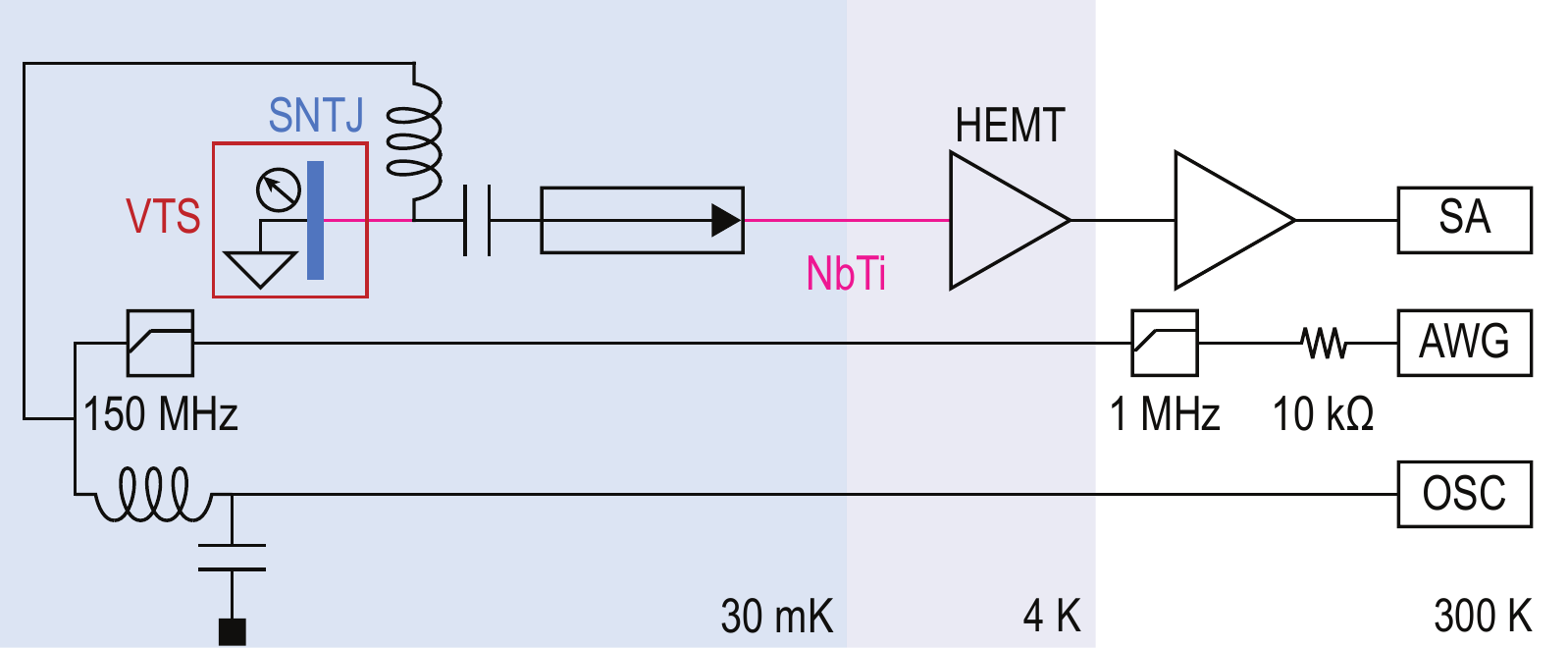}
	\caption{Amplification chain where the HEMT is the first amplifier, placed at 4\,K. The SNTJ (and its packaging) is mounted on a VTS.} 
    \label{fig:hemtalonesetup}
    \end{figure}

	We present in Fig.\,\ref{fig:hemtalonesetup} an amplification chain, where the 4\,K HEMT is the first amplifier. Compared to that of Fig.\,\ref{fig:fullsetup}, several components have been removed from the signal path, notably the bias tees used to deliver the dc current $I_d$ to the KI-TWPA, and the low-pass filter that protects the HEMT from the strong KI-TWPA rf pump tone. We chose to keep the isolator, because it is often placed before the HEMT in qubit experiments to avoid back-action \cite{rosenthal2021efficient} and in satellite mission concepts \cite{Bandler2019Lynx,Echternach2021large}, but we placed it at millikelvin temperatures to minimize its noisy effect. Remaining in the chain are the unavoidable SNTJ packaging and bias tee.
	
\begin{figure*}[t!] 
	\centering
	\includegraphics[scale=0.6]{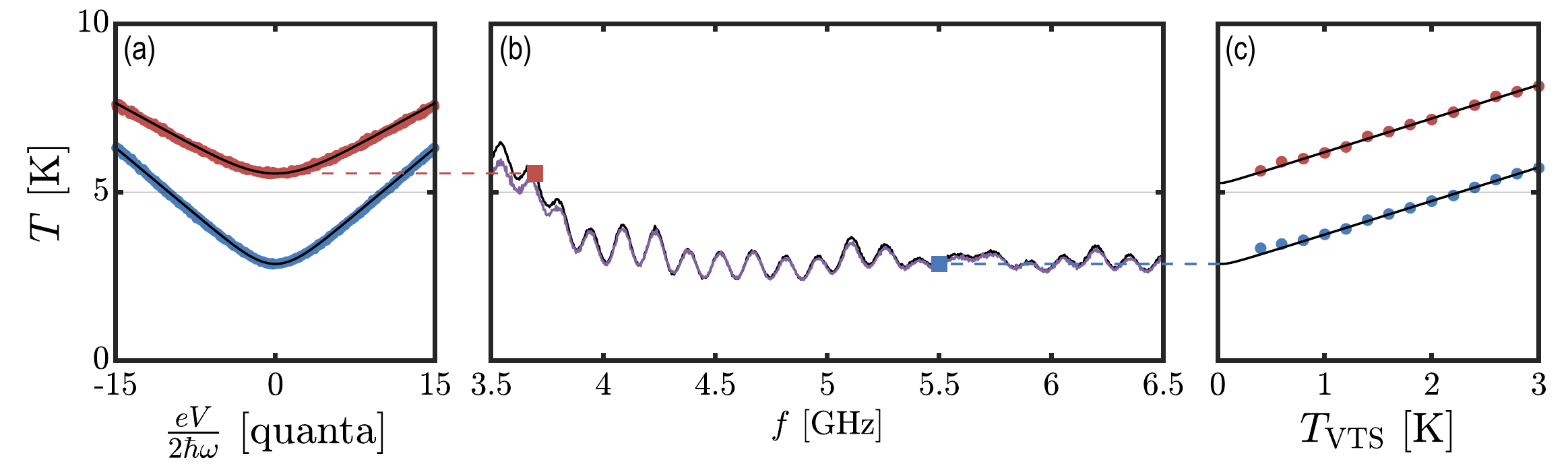}
	\caption{Characterization of a HEMT-only amplifier chain (mounted at 4\,K). (a) The output noise temperature is referred to the chain's input (i.e. divided by the chain's gain), and varies as a function of the SNTJ dc-bias voltage $V$. We illustrate the output noise recorded on the SA, in a 5\,MHz window around two frequencies: 3.6\,GHz (red curve) and 5.5\,GHz (blue curve). We then retrieved the chain-added noise temperature $T_{\Sigma2}$ from a fit (black curves). (b) Two techniques allow us to measure $T_{\Sigma2}$ as a function of frequency: one using the shot noise generated by the dc-biased SNTJ (black curve), and one using the temperature-dependent Johnson noise generated by the unbiased SNTJ, mounted on a VTS (purple curve). Two squares underline the values of $T_{\Sigma2}$: at 3.6\,GHz, obtained with the shot noise method (red) and at 5.5\,GHz, obtained with the Johnson noise method (blue). (c) The input-referred output noise temperature varies as a function of the VTS temperature. We illustrate this variation for two frequencies, 3.6\,GHz (red points) and 5.5\,GHz (blue points). We then recover $T_{\Sigma2}$ from a fit (black lines).}
    \label{fig:HEMTalone}
    \end{figure*}

	Generating shot noise with the dc-biased SNTJ and fitting the output noise recorded on the SA [see Fig.\,\ref{fig:HEMTalone}(a)], we obtain $T_{\Sigma2}$ as a function of frequency, shown in Fig.\,\ref{fig:HEMTalone}(b). To validate this result, we employed another, independent technique: in fact, at the chain's input the SNTJ (with its packaging) is mounted on a variable temperature stage (VTS), allowing us to generate a temperature-dependent Johnson noise with the unbiased SNTJ. Fitting the output noise [see Fig.\,\ref{fig:HEMTalone}(c)] we obtain a second estimate of $T_{\Sigma2}$. Here, the chain's reference plane advances to the SNTJ packaging output, because the packaging's temperature also varies. Then, comparing the chain's gains between the two measurements (SNTJ and VTS), we estimate the SNTJ packaging insertion loss to $0.3\pm0.3$\,dB in the band of interest (see appendix\,\ref{app:lossbudg}), similar to previous evaluations \cite{chang2016noise}. The quantitative agreement between both methods validate our use of the SNTJ as a calibrated noise source.
    
	
	\section{Fit of noise curves}
	
	\subsection{Shot noise curves}
	\label{app:fitshot}
	
	We follow the same fitting procedure than in Ref.\,\cite{malnou2021three}, but in the simpler case where we don't include the idler noise contribution. In short, the SNTJ delivers noise to the chain \cite{lecocq2017nonreciprocal,malnou2021three}, whose power is
	\begin{equation}
	\begin{aligned}
    N_\mathrm{in}^s = \frac{k_B T}{2 \hbar\omega_s}\bigg[&\frac{eV+\hbar\omega_s}{2k_B      T}\coth\left(\frac{eV+\hbar\omega_s}{2k_B T}\right)\\
	+ &\frac{eV-\hbar\omega_s}{2k_B T}\coth\left(\frac{eV-\hbar\omega_s}{2k_B T}\right)\bigg],
	\end{aligned}
	\label{eq:Ni}
	\end{equation}
    where $T$ is the SNTJ temperature, $V$ is the voltage across the SNTJ, and $\omega_s$ the signal frequency. The output noise recorded on the SA is proportional to that of Eq.\,\ref{eq:N4sp}.
    
    In practice, we first fit the output noise at high voltage $V$, for $\lvert eV/(2\hbar\omega_s)\rvert > 3$ quanta. In that case, Eq.\,\ref{eq:Ni} reduces to
    \begin{equation}
	N_\mathrm{in}^s =\frac{eV}{2\hbar\omega_s}.
	\end{equation}
    We obtain the chain's gain $G_c$ and first estimation of $N_\Sigma'$. 
    
    Then we fit the central region, with $G_c$ fixed. We let $N_\Sigma'$ vary within $\pm25\%$ of its first estimation, and because the AWG has a slight voltage offset $V_\mathrm{off}$, we include it as a fit parameter: we write $V-V_\mathrm{off}$ instead of $V$ in Eq.\,\ref{eq:Ni}. Finally, we bound $T$ to a maximum value of $1\,$K.
	
	\subsection{Johnson noise curves}
	\label{app:fitjohnson}
	
	When varying the VTS temperature, we deliver noise to the chain whose power is
	\begin{equation}
	    N_\mathrm{in}^s = \frac{1}{2}\coth\left({\frac{\hbar\omega}{2k_BT_\mathrm{VTS}}}\right).
	\end{equation}
	Knowing $T_\mathrm{VTS}$, we fit the output noise recorded on the SA to get $G_{c2}$ and $N_{\Sigma2}$, respectively the gain and added noise of the chain containing only the HEMT at 4\,K, see in appendix\,\ref{app:4KHEMT}.
	
	\section{Loss budget}
	\label{app:lossbudg}
	
\begin{figure}[!h]
	\centering
	\includegraphics[scale=0.47]{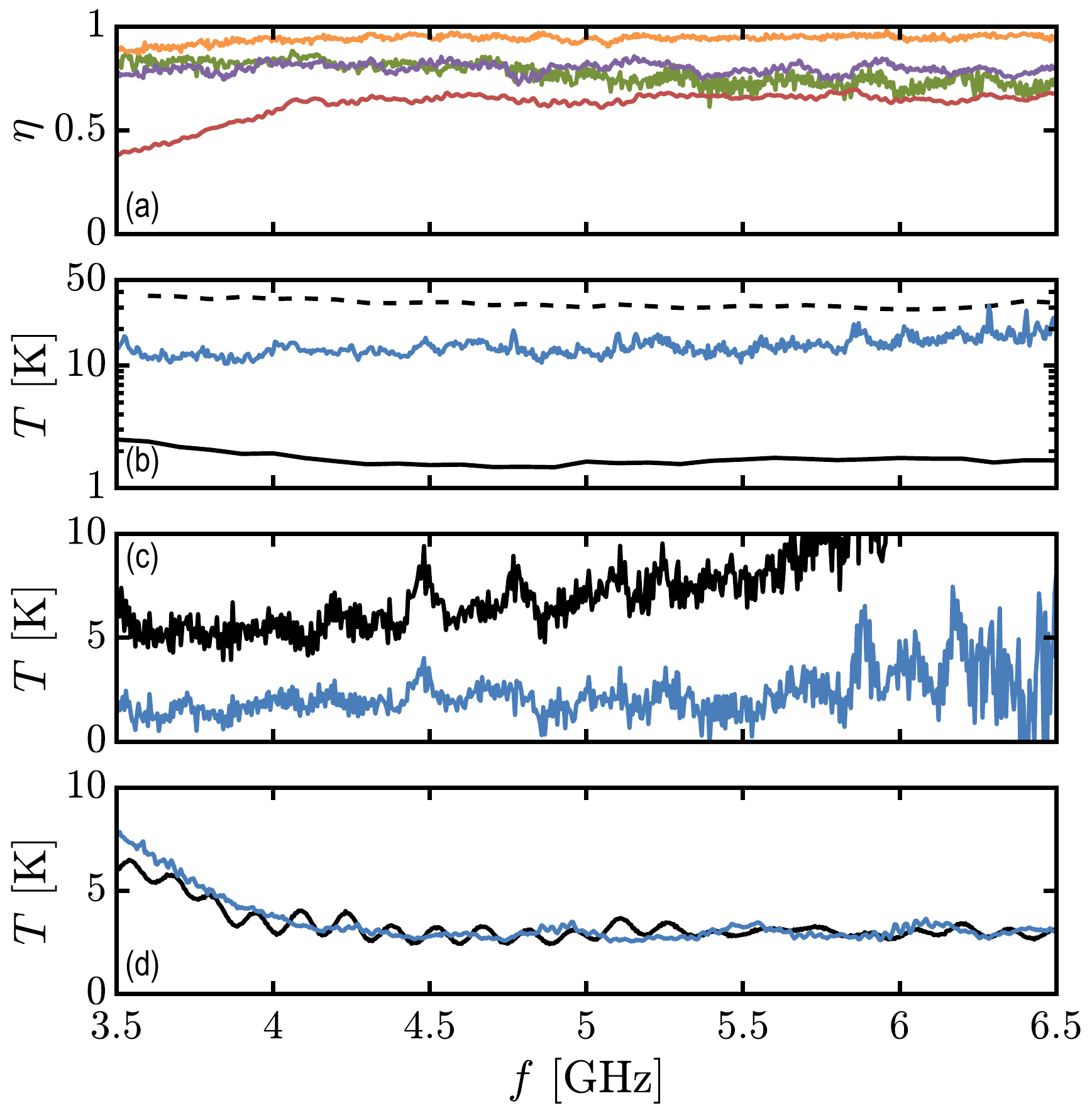}
	\caption{Loss budget and inferred noise temperatures. (a) The transmission efficiencies $\eta_\mathrm{1c}$ (purple line), $\eta_\mathrm{1h}$ (green), $\eta_2$ (red) have been measured at 4\,K. The SNTJ packaging transmission $\eta_p$ (orange line) has been deduced from the ratio of chain's gains, obtained when using the SNTJ and the VTS. (b) The HEMT-added noise temperature $T_H$ (blue line) is then deduced using Eq.\,\ref{eq:Nsigmap}. In comparison, we also show the noise temperature of the HEMT when it is at 4\,K (solid black line) and 296\,K (dashed black line). (c) The KI-TWPA-excess noise temperature $T_\mathrm{ex}$ (blue line) is then calculated with Eq.\,\ref{eq:fullNsig} using the data presented in (a) and (b), and using $T_\Sigma$ (black line). (d) Both the measured (black line) and inferred (blue line) chain-added noise temperature $T_{\Sigma2}$ agree quantitatively well. The inference is made from the constructor specification of the HEMT-added noise at 4\,K and from transmission efficiency measurements on the parts of the chain shown in Fig.\,\ref{fig:hemtalonesetup}.} 
    \label{fig:lossbudg}
    \end{figure}

	We measured at $4$\,K the transmission of each parts of the amplification chain with a VNA, to retrieve $\eta_\mathrm{1c}$, $\eta_\mathrm{1h}$ and $\eta_\mathrm{2}$ as a function of frequency see Fig.\,\ref{fig:lossbudg}(a). To find $\eta_\mathrm{1c}$, we first measured the transmission $\eta_\mathrm{\alpha c}^s$ of the chain from the output of the SNTJ packaging to the input of the NbTi cable that connects the 30\,mK stage to the 4\,K stage. The SNTJ being a one-port device, we cannot measure directly the transmission of the SNTJ packaging $\eta_p$. Instead, we find it from the noise measurements performed on the chain solely containing the HEMT at 4\,K: $\eta_p=0.93\pm0.07$ is the ratio between the chain's gain obtained when using the SNTJ (proportional to $G_H\eta_2\eta_\mathrm{1h}\eta_\mathrm{\alpha c}^s\eta_p$) to the gain obtained when using the VTS (proportional to $G_H\eta_2\eta_\mathrm{1h}\eta_\mathrm{\alpha c}^s$). We then have $\eta_\mathrm{1c} = \eta_\mathrm{\alpha c}^s \eta_p$. The KI-TWPA packaging transmission is measured separately at 4\,K with a low probe tone power. It is then equally divided between $\eta_\mathrm{1h}$ and $\eta_2$.
	
	Knowing all the transmission efficiencies, we use Eq.\,\ref{eq:Nsigmap} to deduce $N_H$, the HEMT-added noise at 70\,K, from $N_\Sigma'$, the noise added by the chain presented in Fig.\,\ref{fig:setup}(c) (and measured with the SNTJ). In Fig.\,\ref{fig:lossbudg}(b) we present $T_H = N_H \hbar\omega/k_B$ as a function of frequency, along with the constructor specifications for the HEMT-added noise, when the HEMT is at 4\,K and 296\,K \cite{lnf}. Unsurprisingly, $T_H$ lies between these two reported performances, with $T_H=13.4\pm0.4$\,K between 3.5 and 5.5\,GHz (uncertainty dominated by that of the transmission efficiencies).
	
	Then, using Eq.\,\ref{eq:fullNsig} we deduce $N_\mathrm{ex}$, the KI-TWPA-excess noise. In Fig.\,\ref{fig:lossbudg}(c) we show $T_\mathrm{ex}=N_\mathrm{ex} \hbar\omega/k_B$, along with $T_\Sigma$, as a function of frequency. In the 3.5-5.5\,GHz band we have $T_\mathrm{ex}=1.9\pm0.2$\,K.
	
	Finally, we also measured (at 4\,K) the transmission efficiencies of each parts of the chain presented in Fig.\,\ref{fig:hemtalonesetup}, where the HEMT is mounted at 4\,K. Then, from the constructor specification of the HEMT-added noise at 4\,K [see Fig.\,\ref{fig:lossbudg}(b)] we calculated the expected chain-added noise, and compared it to the measured one, see Fig.\,\ref{fig:lossbudg}(d). Both are in good quantitative agreement, validating our overall methodology for the loss budget.
	
	\section{Four-point probe setup}
	\label{app:pow}
	
	\begin{figure}[!h]
	\centering
	\includegraphics[scale=0.53]{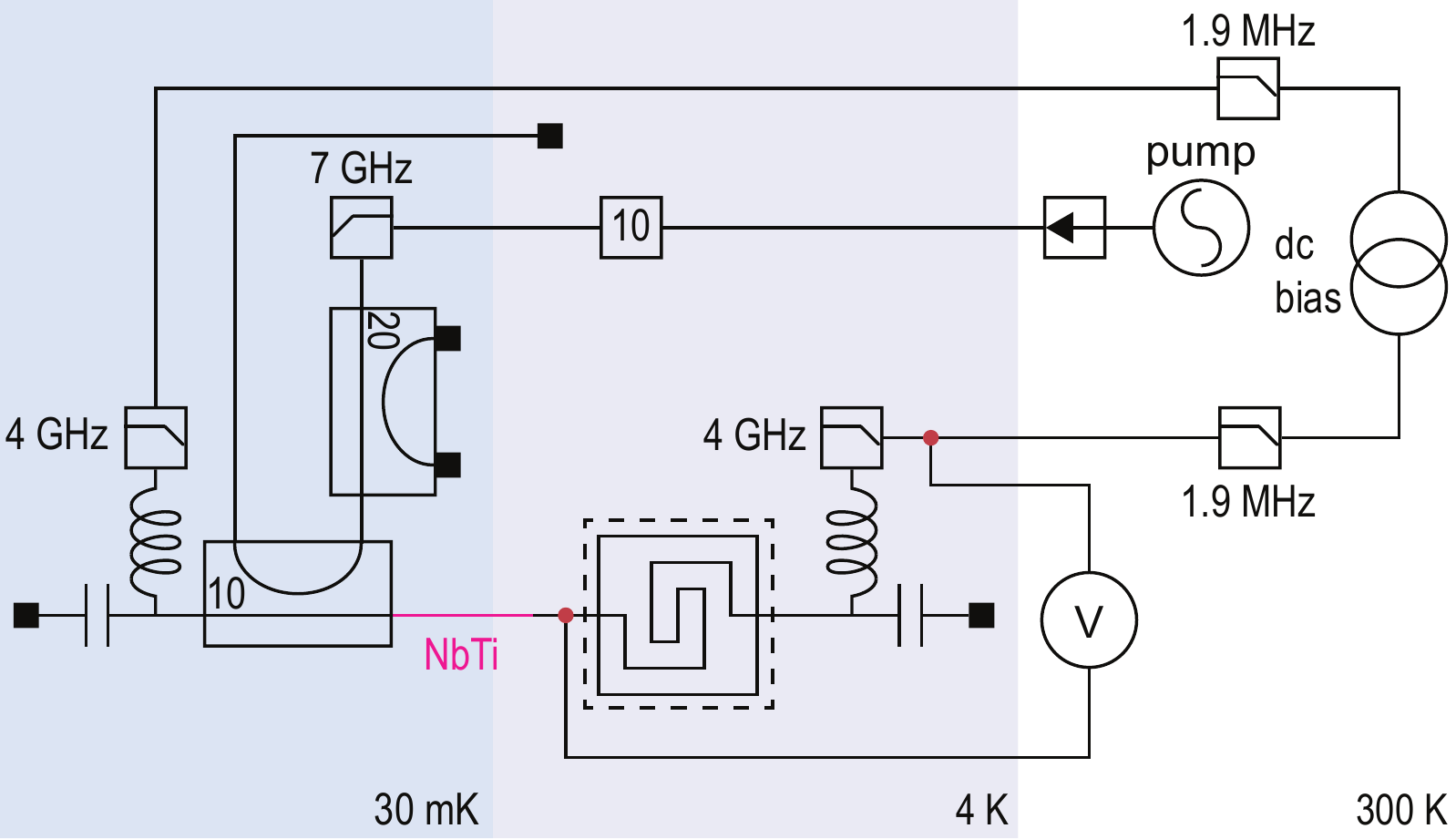}
	\caption{Four-point probe experimental setup. Two red points highlight the ports across which we measure the voltage drop while the KI-TWPA is under dc and rf biases.} 
    \label{fig:powsetup}
    \end{figure}
	
	We present in Fig.\,\ref{fig:powsetup} the experimental setup used to evaluate the dc power consumption of the KI-TWPA. With the KI-TWPA under dc and rf biases, we measure the voltage drop between the 4\,K mandatory components used to operate the KI-TWPA, consisting of the KI-TWPA itself, the BT, and the LPF placed after the dc input of the BT. At room temperature, a voltmeter reads the voltage across these components. Note that we have terminated the rf amplification chain at 4\,K, because it is not properly matched to $50\,\Omega$ anymore: in fact, we inserted a subminiature version A (SMA) T-junction at the KI-TWPA input in order to read the potential at this point.

	\vspace{0.1in}
	
	%

\end{document}